\documentclass[aps,prb,twocolumn,showpacs,superscriptaddress]{revtex4-1}
\usepackage{graphicx}
\begin{document}

\title{Silicite: the layered allotrope of silicon}

\author{Seymur Cahangirov}
\affiliation{Nano-Bio Spectroscopy Group and ETSF Scientific Development Centre, Departamento de F\' isica de Materiales, Universidad del Pa\' is Vasco, CSIC-UPV/EHU-MPC and DIPC, Avenida de Tolosa 72, E-20018 San Sebastian, Spain}

\author{V. Ongun \"{O}z\c{c}elik}
\affiliation{UNAM-National Nanotechnology Research Center, Bilkent University, 06800 Ankara, Turkey}
\affiliation{Institute of Materials Science and Nanotechnology, Bilkent University, Ankara 06800, Turkey}

\author{Angel Rubio}\email{angel.rubio@ehu.es}
\affiliation{Nano-Bio Spectroscopy Group and ETSF Scientific Development Centre, Departamento de F\' isica de Materiales, Universidad del Pa\' is Vasco, CSIC-UPV/EHU-MPC and DIPC, Avenida de Tolosa 72, E-20018 San Sebastian, Spain}

\author{Salim Ciraci}\email{ciraci@fen.bilkent.edu.tr}
\affiliation{Department of Physics, Bilkent University, Ankara 06800, Turkey}

\date{\today}

\begin{abstract}
Based on first-principles calculation we predict two new thermodynamically stable layered-phases of silicon, named as silicites, which exhibit strong directionality in the electronic and structural properties. As compared to silicon crystal, they have wider indirect band gaps but also increased absorption in the visible range making them more interesting for photovoltaic applications. These stable phases consist of intriguing stacking of dumbbell patterned silicene layers having trigonal structure with $\sqrt{3} \times \sqrt{3}$ periodicity of silicene and have cohesive energies smaller but comparable to that of the cubic diamond silicon. Our findings also provide atomic scale mechanisms for the growth of multilayer silicene as well as silicites.
\end{abstract}


\maketitle

\section{Introduction}

Early studies \cite{takeda,engin} to answer the critical question of whether Si can form graphene-like monolayer structures have been ruled out initially by the arguments that Si does not have a layered allotrope like graphite \cite{pauling,rubio,cohen}. Actually, the realization of single and multilayer Si would be rather tempting, since the adaptation of the formidable material and device technology developed for Si crystal to Si nanostructures and its layered compounds is rendered possible.

Recently, the silicon counterpart of graphene named silicene was shown to be stable \cite{seymur1} and was synthesized on Ag substrate \cite{lelay1}. Similar to graphene, the electrons of silicene behave as massless Dirac fermions and armchair silicene nanoribbons display family behavior \cite{seymur1,seymur2}. Much recent growth of multilayer silicene up to $\sim$100 layers \cite{hundred} showing $\sqrt{3} \times \sqrt{3}$ pattern rekindled the fundamental question whether silicon can have stable, graphite-like layered phase \cite{feng,chen,lelay3,vogt}.

Here, by first-principles calculations we show that the structural transformations through interlayer atom transfer results in complex stacking sequence of grown layers, which, eventually can make thermodynamically stable, bulk layered allotropes of silicon. We have named these materials as silicites inspired by the name of layered bulk structure of carbon, graphite. Our predictions herald that the missing layered bulk phases of silicon can, in fact, be synthesized and can add novel properties to those of Si crystal in cubic diamond structure (cdSi), the global minimum of Si. Our findings also provide for a plausible growth mechanism of multilayer silicene, as well as germanene, SiGe and SiC.

\section{Methods}

We have performed state-of-the-art density functional theory (DFT) calculations within generalized gradient approximation(GGA). We used projector-augmented wave potentials PAW \cite{blochl94} and the exchange-correlation potential is approximated with Perdew-Burke-Ernzerhof, PBE functional \cite{pbe}. The Brillouin zone was sampled by (12$\times$12$\times$12) \textbf{k}-points in the Monkhorst-Pack scheme where the convergence in energy as a function of the number of \textbf{k}-points was tested. The number of \textbf{k}-points were further increased to (48$\times$48$\times$48) in DOS calculations. In the RPA calculation we have used (24$\times$24$\times$24) and (88$\times$88$\times$88) \textbf{k}-point mesh for eLDS and cdSi respectively. G$_0$W$_0$ calculation was performed using (8$\times$8$\times$8) \textbf{k}-point mesh and 288 bands whereby convergence with respect to all parameters is ensured for eLDS \cite{shishkin}. Also hybrid functional calculations were carried out for energy band structure \cite{HSE06}. Atomic positions were optimized using the conjugate gradient method. The energy convergence value between two consecutive steps was chosen as $10^{-5}$ eV. A maximum force of 0.002 eV/\AA~ was allowed on each atom. Phonon dispersions are calculated using small displacement method \cite{alfe} where forces are obtained using the VASP software \cite{vasp}. In the molecular dynamics simulations atomic velocities are scaled each 50 steps corresponding to 0.1 ps.

\section{Dumbbell Silicene}

The dumbbell (DB) structure \cite{can,kaltsas,ongun1,ongun2} of silicene has been the most critical ingredient in the construction of the layered bulk phase of Si. A single Si adatom initially at the top site of silicene pushes down the parent Si atom underneath to arrange a DB \cite{ongun1} structure as shown in Fig.~1(a). Hence, a single, isolated DB formation on silicene is an exothermic process and takes place spontaneously \cite{ongun1,ongun2}.

The synthesis of layered phase of Si starts by the growth 3$\times$3 silicene, which is lattice matched to (4$\times$4) Ag(111) substrate. Due to significant interaction between Si and Ag, this commensurate growth on Ag substrate is found to be more favorable energetically than the growth of cdSi \cite{ag111}. Once grown, the 3$\times$3 silicene transforms gradually through chemisorption of Si adatoms forming DBs and eventually displays $\sqrt{3} \times \sqrt{3}$ - R(30$^o$) pattern. At the end, it is shrank by $\sim$5\% and hence it becomes incommensurate with Ag(111) surface \cite{feng,chen}. 

It has been recently identified that this modified monolayer grown on Ag(111) surface is honeycomb dumbbell structure (HDS) and it comprises two DBs in each $\sqrt{3} \times \sqrt{3}$ cell, which are arranged in a honeycomb structure as shown in Fig.~1(b) \cite{seymur4}. The lattice constant of the $\sqrt{3} \times \sqrt{3}$ silicene and the stepheight between 3$\times$3 and $\sqrt{3} \times \sqrt{3}$ silicene reported experimentally are well reproduced by the growth mechanism mentioned above \cite{chen, lelay3}. Free standing HDS (with energy 0.912 eV/\AA$^2$), as well as HDS on Ag substrate (with energy 1.014 eV/\AA$^2$) are found to be stable \cite{fds}. Small difference of energies of HDSs with and without substrate of 0.1 eV/\AA$^2$, which becomes even smaller in real incommensurate substrate, is a strong evidence that the effect of substrate diminish in multilayer structure.

Upon the growth of a second monolayer, first the grown HDS is forced to change to TDS (i.e. the trigonal dumbbell structure, which has only one DB in the $\sqrt{3} \times \sqrt{3}$ cell and is less energetic then HDS). As shown in Fig.~1(c), the upper Si atom of every other DB in each unit cell of existing HDS is transferred to the upper monolayer to form a DB above. Then one Si atom of each DB at either layer is connected to a threefold coordinated regular Si atom and hence forms a perpendicular Si-Si bond recovering the fourfold coordination. At the end, both the second grown layer and the existing HDS change into TDS. However, once the second TDS completed, it changes eventually to HDS through adsorption of incoming Si adatoms in the medium, because it is energetically favorable. Hence, these structural transformations occur without the need to overcome any kind of energy barrier and they follow sequentially whenever a new Si monolayer grows on top. At the end, all grown layers change from HDS to TDS one by one, except the last grown layer which remains to be HDS. This growth mechanism matches very well the measured STM and LEED data \cite{lelay3,feng,chen,vogt,seymur4}. For example, observation of honeycomb pattern in STM images obtained from last grown layers silicene multilayers is in compliance with our results, since last grown layer has to be HDS and calculated STM images of HDS display the same honeycomb pattern.

\begin{figure}
\includegraphics[width=8.5cm]{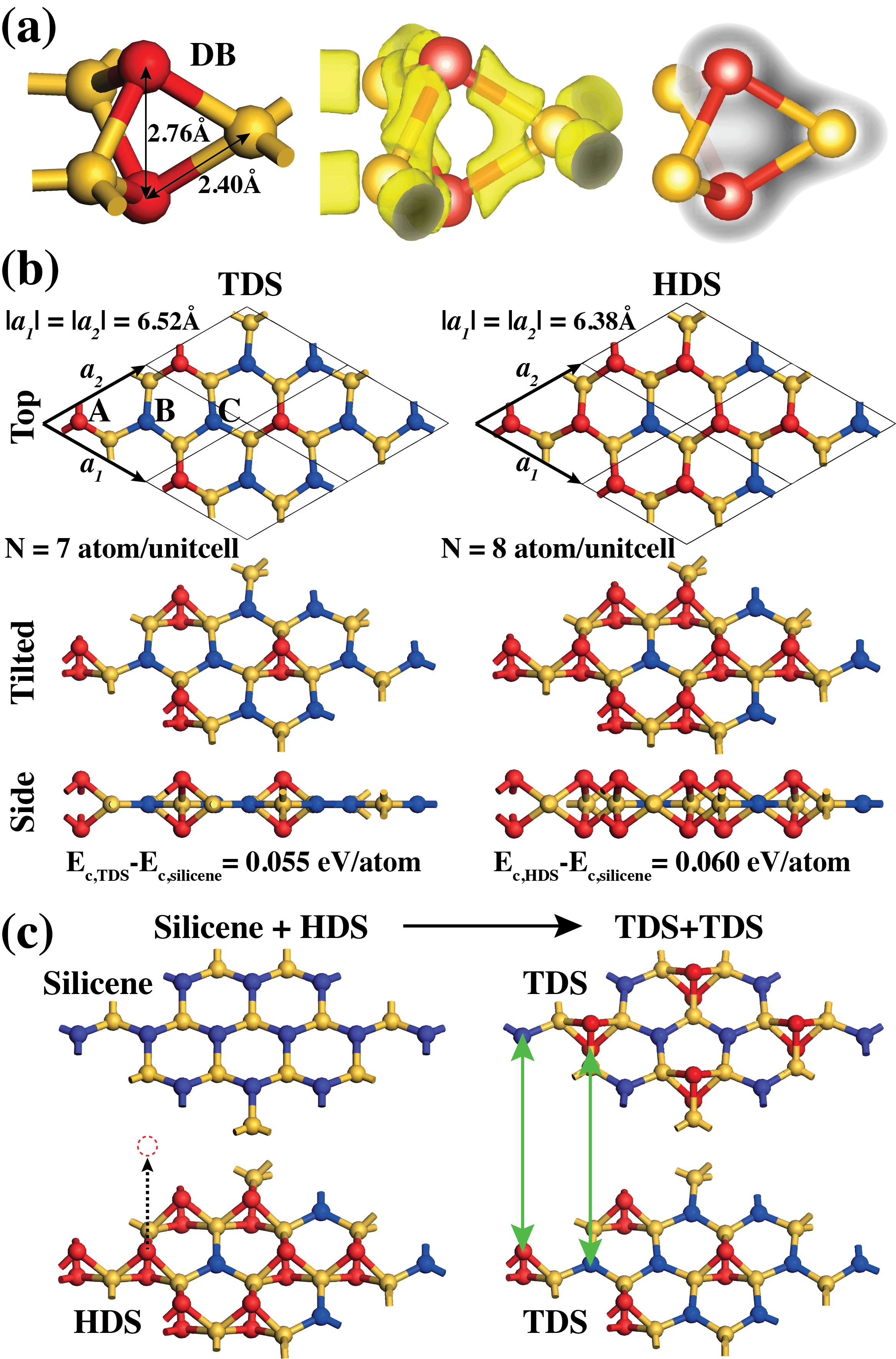}
\caption{(a) Single dumbbell (DB) structure together with isosurface and contour plots of the total charge density. (b) Top and side views of trigonal dumbbell structure (TDS) and honeycomb dumbbell structure (HDS) having $\sqrt{3} \times \sqrt{3}$ - R(30$^o$) pattern. Red, yellow and blue balls represent dumbbell, fourfold and threefold coordinated Si atoms of TDS, respectively. A, B and C are three stacking positions. In HDS yellow Si atoms are fivefold coordinated. (c) Growth mechanism through the structural transformation by the transfer of Si atom  from the lower HDS to the upper, newly grown silicene forming eventually two TDSs are indicated by dotted arrow and subsequent Si-Si rebounding are indicated by green arrows. The upper TDS is shifted to the B-position of the lower TDS.}
\label{fig1}
\end{figure}

\section{Layered Dumbbell Silicites}

The structural transformations in grown layers and the resulting stacking sequence summarized above have led us to predict two different layered bulk phases of silicon; we named as silicite. The atomic structures of these phases are presented in Fig.~2. In both phases, top and bottom Si atoms of a DB in each unit cell of TDS are bonded to the regular, threefold coordinated Si atoms at either side through two perpendicular Si-Si bonds. At the end, the third site in the unit cell becomes bonded to the lower adjacent TDS only by leaving a hole below the upper adjacent TDS. Accordingly, the fourfold coordination of all Si atoms is maintained in these layered dumbbell silicite (LDS) structures, but the bond angles deviate significantly from tetrahedral angle, $\sim$109$^o$. In the first phase, named here as eclipsed layered dumbbell silicite (eLDS), the layers follow an ABCABC... stacking sequence as depicted in the top view of TDS in Fig.~1(b). Three in-layer Si-Si bonds in either side of the perpendicular Si-Si bonds attain similar orientations as in the left panel of Fig.~2(a). Consequently, all in-layer Si-Si bonds in different TDSs are nearly eclipsed, so that the top view of the multilayer looks like a single layer honeycomb. In the other phase, namely staggered layered dumbbell silicite (sLDS) the stacking sequence is A$\bar{B}$C$\bar{A}$B$\bar{C}$A... The in-layer Si-Si bonds of the upper and lower TDS layers oozing from both ends of perpendicular Si-Si bonds specified by the sign "bar" are staggered by 60$^o$ as shown by the inset in the right panel of Fig.~2(a). eLDS and sLDS in the direction perpendicular to TDS layers yield views of atoms reminiscent of the view along [111] direction of eclipsed and staggered cdSi, respectively. The top views of these layered silicites are compared with diamond silicon structures seen along [111] direction in Fig.~2. Apparently, the stacking of LDS structures that we predict here rules out earlier presumed multilayer of pristine silicene having a Bernal stacking like graphite, since the latter is not energetically favorable and cannot match LEED patterns.

\begin{figure}
\includegraphics[width=8.5cm]{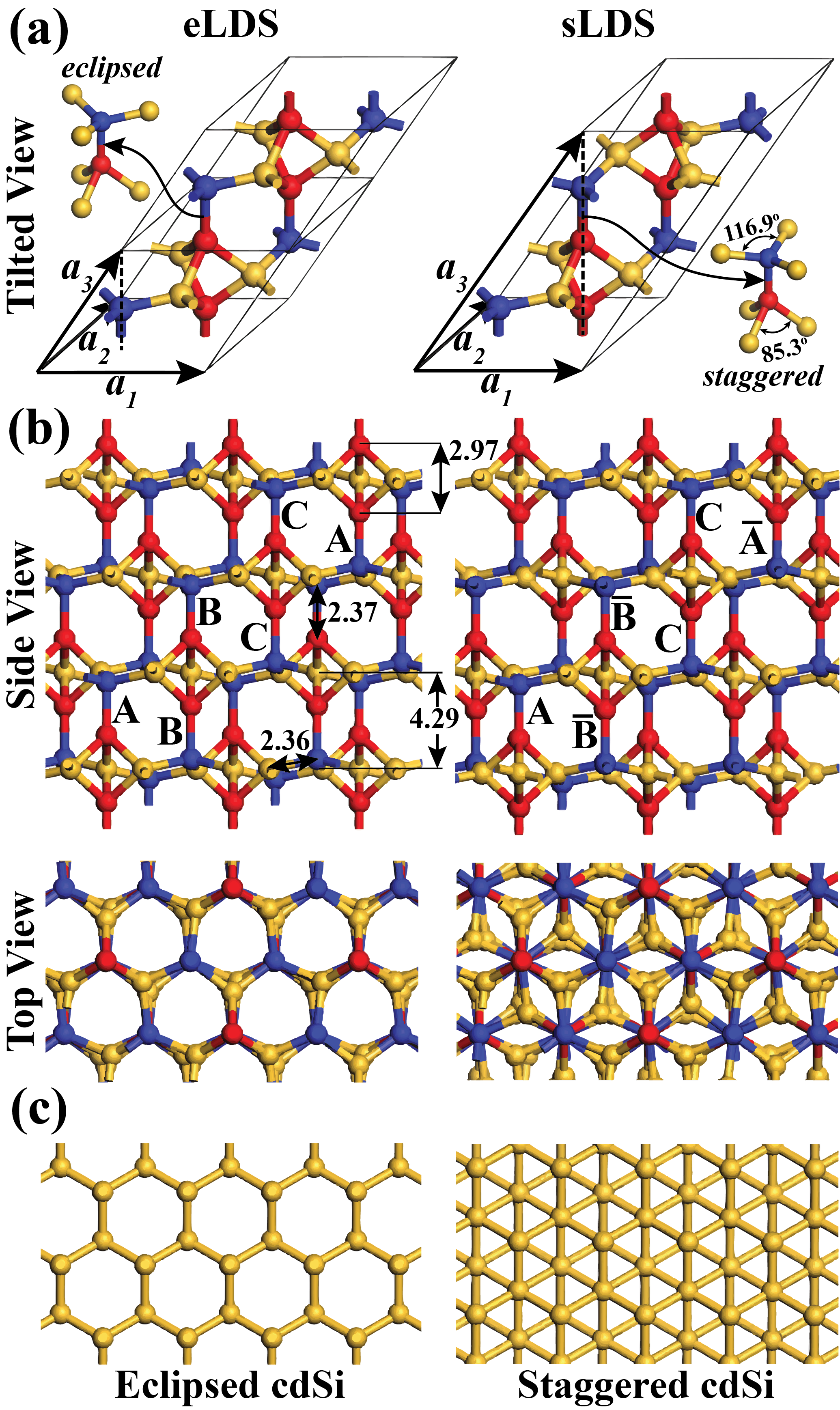}
\caption{(a) The double unit cell of eclipsed layered dumbbell silicite (eLDS) including $N$=7 Si atoms per unit cell and single unit cell of staggered layered dumbbell silicite (sLDS) including $N$=14 Si atoms per unit cell. (b) Side and top views showing the ABCABC... stacking of eLDS and the A$\bar{B}$C$\bar{A}$B$\bar{C}$A... stacking of sLDS. (c) Top view of eclipsed and staggered diamond structure of silicon is shown for comparison.}
\label{fig2}
\end{figure}

Notably, the unit cell of sLDS is twice larger than that of eLDS and comprises 14 Si atoms. Both eLDS and sLDS are viewed as layered materials, since they consist of parallel TDS layers; in each TDS layer DBs show $\sqrt{3} \times \sqrt{3}$ pattern. Since there is only one hole between two consecutive layers in the unit cell, the mass densities of eLDS (2.10 g/cm$^3$) and of sLDS (2.11 g/cm$^3$) are slightly smaller than that of cdSi (2.28 g/cm$^3$). We note that in cdSi the distance between (111) planes is only 2.37 \AA. On the other hand, because of two perpendicular Si-Si chemical bonds in each unit cell which connect the adjacent TDS layers, the interaction between parallel layers of LDS structures is not like the weak van der Waals interaction in graphite or MoS$_2$ \cite{can1}. The calculated cohesive energies (i.e. the total energy per atom relative to the energy of free Si atom) are 4.42 eV and 4.43 eV for eLDS and sLDS, respectively, which are only 0.18 - 0.17 eV smaller than that of cdSi while 0.46 - 0.47 eV larger than that of freestanding silicene \cite{vdw}.

The calculated vibrational frequencies of eLDS and sLDS phases are all found to be positive. The absence of negative frequencies is taken as an evidence that these layered phases are stable. The phonon bands of these structures presented in Fig.~3 disclose interesting dimensionality effects. Specific optical branches are flat and the lower lying branches overlap with the acoustical branches. For sLDS structure the acoustical branch dips in the $\Gamma$-L direction leading to phonon softening. We have also performed molecular dynamics simulations where (3$\times$3$\times$4) supercell of eLDS and (3$\times$3$\times$2) supercell of sLDS were kept at 1000 K for 4 ps. No structural deformation was observed in the course of these simulations, which corroborates the stability of these materials.

\begin{figure}
\includegraphics[width=8.5cm]{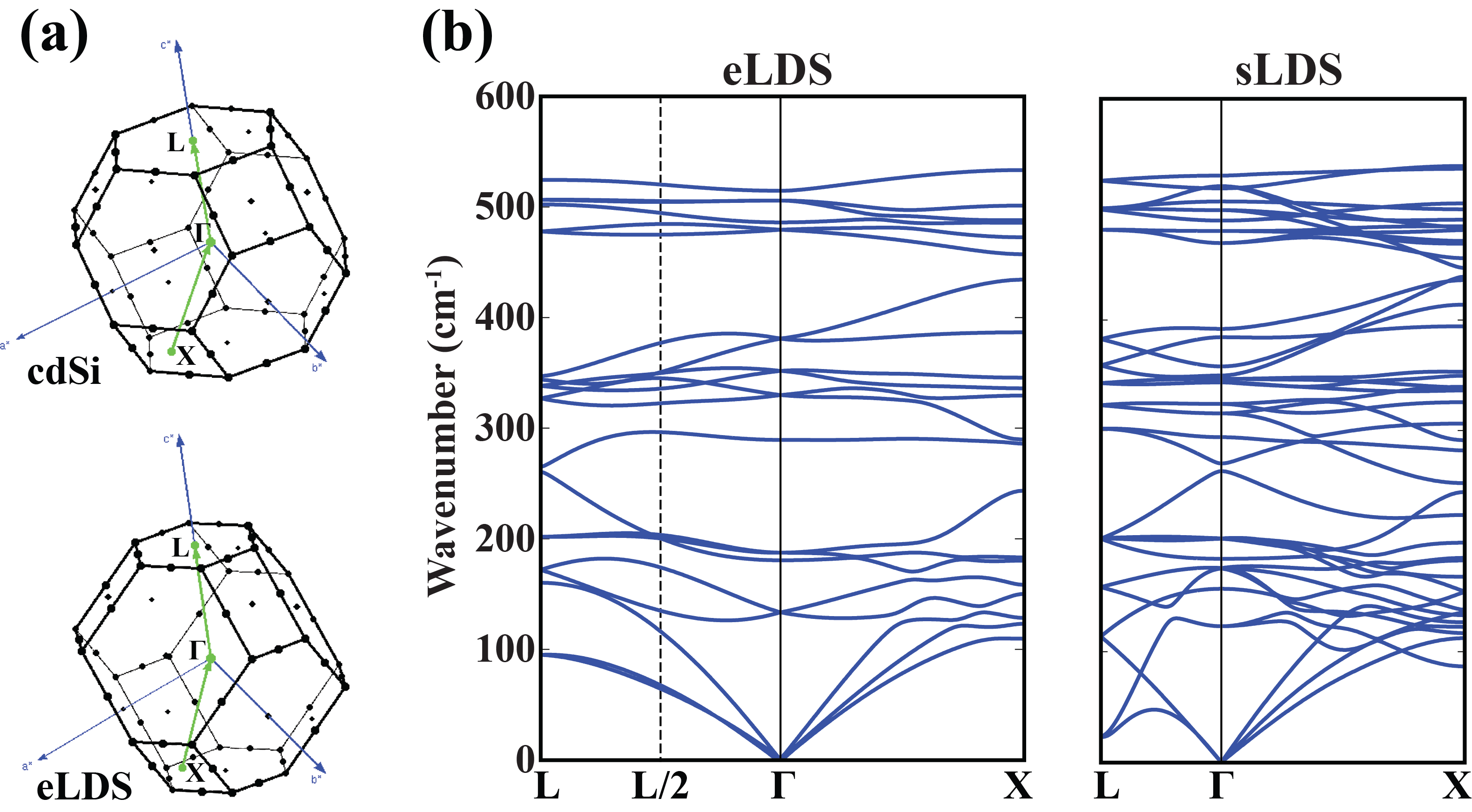}
\caption{(a) Brillouin Zones of cdSi and eLDS with relevant symmetry directions. (b) Phonon bands of eLDS and sLDS.}
\label{fig3}
\end{figure}

\begin{figure}
\includegraphics[width=7.9cm]{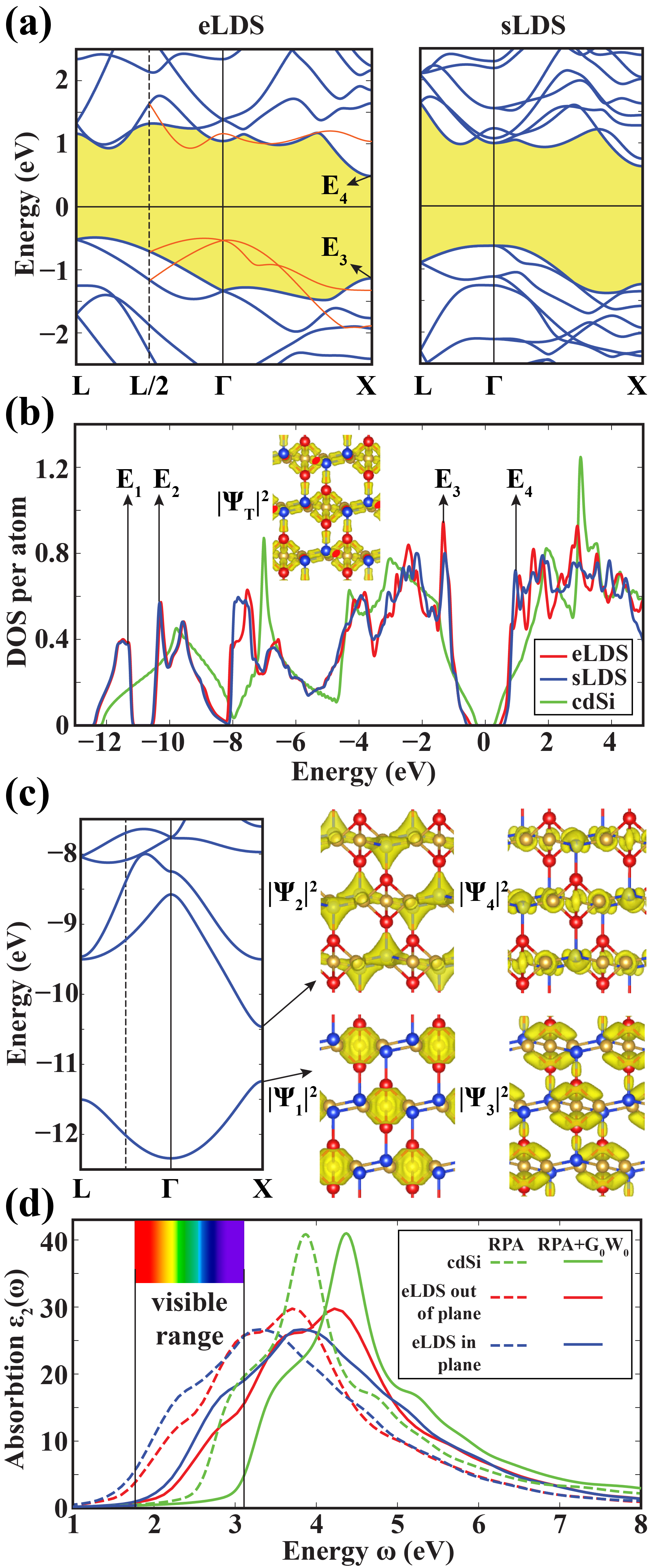}
\caption{(a) Energy band structure of eLDS and sLDS along L-$\Gamma$-X directions of the Brillouin zone. Zero of energy is set to the the Fermi level. Bands of eLDS folded by doubling the unit cell along $a_{3}$ are shown by red lines. (b) Normalized densities of states (DOS) of eLDS, sLDS and cdSi. The isosurfaces of the total charge density shown by inset confirm the layered nature. (c) Energy band structure of eLDS around the gap in the valence band together with the isosurface charge density of the states $\Psi_{1-4}$ leading to the peaks, $E_{1-4}$ in DOS. (d) The calculated Kohn-Sham and G$_0$W$_0$ RPA optical absorption spectra for eLDS and cdSi.}
\label{fig4}
\end{figure}

Whether eLDS and sLDS carry the characteristic features of a layered material can be conveniently substantiated by investigating the in-plane and out of plane Young modulus and by comparing them with those of cdSi. Perpendicular Young's modulus of eLDS and sLDS are calculated as $Y_{\perp}$=79.6 GPa and 76.4 GPa, respectively,  while the Young's modulus of cdSi along [111] direction is 176.0 GPa and hence more than twice the value of LDS phases. In contrast, the in-plane Young's modulus calculated within  TDS layers of eLDS and sLDS are relatively higher, and are 176.3 GPa and 161.9 GPa, respectively. These values are comparable with the Young's modulus of cdSi calculated in the (111) plane, which is 200 GPa. The dramatic differences between the Young's modulus of LDS structures and cdSi calculated in the direction perpendicular to layers confirm the layered nature of LDS phases.

The electronic structure of the layered phases have indirect band gaps, which are wider than that of cdSi, as shown in Fig.~ 4(a). The calculated indirect (direct) band gaps of eLDS and sLDS are 0.98 (1.43) eV and 1.26 (1.65) eV, respectively. The indirect band gap of cdSi is 0.62 eV at the DFT-PBE level while it is increased to 1.12 eV upon including many-body self-energy corrections at the G$_0$W$_0$ level \cite{shishkin}. With G$_0$W$_0$ correction the indirect band gap of eLDS increased to 1.52 eV. Indirect (direct) band gaps of eLDS and sLDS calculated by HSE06 hybrid functional are 1.92 eV (2.37 eV) and 1.88 eV (2.26 eV), respectively.

Owing to the different Brillouin zones it is difficult to directly compare the band structures of LDS and cdSi. Therefore the effects of the layered character on the electronic structure are sought in the normalized densities of states (DOS). Figure~4(b) shows the normalized DOSs of eLDS, sLDS and cdSi. Except for some shifts of peaks, DOSs of silicites are similar. Owing to the fourfold coordination of Si atoms in all structures, the overall features of DOSs of LDS structures appear to be reminiscent of that of cdSi. This confirms the fact that the overall features of the bands of cdSi can be obtained within the first nearest neighbor coupling \cite{harrison}. The total charge density, $|\Psi_T|^2$ presented by inset, depicts that electrons are mainly confined to TDS layers. This is another clear manifestation of the layered character of eLDS and sLDS phases. On the other hand, significant differences are distinguished in the details of electronic energy structure due to deviations from tetrahedral coordination: (i) Indirect band gaps relatively larger than that of cdSi can offer promising applications in micro and nanoelectronics. (ii) Sharp peaks $E_3$ and $E_4$ near the edges of the valence and conduction bands, originate from the states, which are confined to TDS layers and can add critical functionalities in optoelectronic properties. (iii) A gap opens near the bottom of the valence band at $\sim$ -11 eV; its edge states are also confined to TDS layers.

In Fig.~4(d) we present the optical absorption spectra of eLDS and cdSi calculated at the RPA level using the Kohn-Sham wave functions and G$_0$W$_0$ corrected eigenvalues. One can see that the optical absorption of eLDS is significantly enhanced in the visible range compared to cdSi which makes it a potential candidate material for photovoltaic applications. This enhancement is still present when we rigidly shift the absorption spectra by the amount we get from G$_0$W$_0$ corrections \cite{revmod}. 

The in-plane and out of plane static dielectric response also reflects the layered nature of silicite. The frequency dependent dielectric matrix takes different values in the in-plane and out of the plane directions of eLDS while for cdSi it is isotropic. The calculated in-plane dielectric constant of eLDS (sLDS) is $\epsilon_{\parallel}$=12.52 (12.85), while its out of plane dielectric constant is $\epsilon_{\perp}$=11.69 (11.56). Those values are contrasted with the uniform dielectric constant, $\epsilon$=12.19 of cdSi.

\section{Conclusion}

In conclusion, we propose the growth mechanism for layered allotropes of silicon with each layer composed of dumbbell configurations arranged in a $\sqrt{3} \times \sqrt{3}$ supercell. Our analysis based on the state-of-the-art first-principles calculations show that these phases are thermodynamically stable and have energies only 0.17-0.18 eV above the global minimum. Their elastic and electronic properties display pronounced directionality. Enhanced absorption in the visible spectrum makes layered phases superior to cdSi for photovoltaic applications. The puzzling STM and LEED data obtained at different stages of multilayers grown on Ag (111) surface are successfully interpreted.

\section{Acknowledgements}

We thank E. Durgun for his assistance in HSE calculations. The computational resources have been provided by TUBITAK ULAKBIM, High Performance and Grid Computing Center (TR-Grid e-Infrastructure). S. C. and V. O. \"{O} acknowledge financial support from the Academy of Sciences of Turkey (TUBA). S. Cahangirov and A. Rubio acknowledges financial support from the Marie Curie grant FP7-PEOPLE-2013-IEF project number 628876, the European Research Council  (ERC-2010-AdG-267374), Spanish Grant (FIS2010-21282-C02-01), Grupos Consolidados (IT578-13), and the EU project (280879-2 CRONOS CP-FP7).


\begin{thebibliography}{35}
\bibitem{ag111}
The cohesive energy of the grown $3\times3$ silicene on Ag substrate is 4.88 eV/atom, which is higher than that of free standing silicene (3.96 eV/atom) as well as cdSi (4.60 eV/atom).
\bibitem{fds}
Even if the energy is gained further by the adsorption of additional single Si atom in the $\sqrt{3} \times \sqrt{3}$ - R(30$^o$) unit cell of HDS to form Full Dumbbell Structure (FDS), the cohesive energy per Si atom is lowered and FDS by itself is found unstable.
\bibitem{vdw}
The difference of cohesive energy between sLDS and eLDS remained to be $\sim$0.1 eV and their lattice constants are reduced by $\sim$1.3\%, when the van der Waals correction is included.
\end{thebibliography}
\end{document}